\documentclass{article}

\usepackage{natbib}
\bibpunct{}{}{,}{s}{}{}

\usepackage{graphicx}
\usepackage{dcolumn}
\usepackage{bm}
\usepackage{upgreek} 
\usepackage{wasysym} 

\begin{document}


\title{Macroscopic graphene membranes and their extraordinary stiffness}

\author{Tim J. Booth \footnote{author to whom correspondence should be addressed: \texttt{tim.j.booth@gmail.com}} \footnote{Manchester University, Department of Physics and Astronomy, Schuster Laboratory, Brunswick Street, Manchester M13 9PL, UK},
Peter Blake\footnote{Graphene Industries Ltd, 32 Holden Avenue, Manchester M16 8TA, UK},
Rahul R. Nair\footnotemark[2], Da Jiang\footnotemark[3],\\ Ernie W. Hill\footnote{Manchester University, Center for Mesoscience and Nanotechnology, Oxford Road, Manchester M13 9PL}, Ursel Bangert\footnote{Manchester University, Materials Science Center, Grosvenor Street, Manchester M1 7HS, UK}, Andrew Bleloch\footnote{SuperSTEM, Daresbury Laboratory, Daresbury, Cheshire WA4 4AD, UK}, Mhairi Gass\footnotemark[6], \\Kostya S. Novoselov\footnotemark[2], M.~I.~Katsnelson\footnote{Institute for Molecules and Materials, Radboud
University Nijmegen, 6525 AJ, Nijmegen, The Netherlands} and A.~K.~Geim\footnotemark[2] }

\maketitle


\begin{abstract}
The properties of suspended graphene are currently attracting enormous interest,
but the small size of available samples and the difficulties in making
them severely restrict the number of experimental techniques that
can be used to study the optical, mechanical, electronic, thermal and
other characteristics of this one-atom-thick material. Here we
describe a new and highly-reliable approach for making graphene
membranes of a macroscopic size (currently up to 100 $\upmu$m in
diameter) and their characterization by transmission electron
microscopy. In particular, we have found that long graphene beams
supported by one side only do not scroll or fold, in striking
contrast to the current perception of graphene as a supple thin
fabric, but demonstrate sufficient stiffness to support extremely
large loads, millions of times exceeding their own weight, in agreement
with the presented theory. Our work opens many avenues for studying
suspended graphene and using it in various micromechanical systems
and electron microscopy.
\end{abstract}



Graphene is a one-atom-thick crystal consisting of carbon atoms that
are $sp^2$-bonded into a honeycomb lattice. Its exceptional
properties continue to attract massive interest, making graphene
currently one of the hottest topics in materials science
\cite{review}. Much experimental work has so far been carried out on
graphene flakes produced on top of oxidized silicon wafers by
micromechanical cleavage \cite{aEFE,aPNAS,zAFMgraphene}. More
recently, procedures were developed to process graphene crystallites
further and obtain suspended (free-standing) graphene
\cite{suspend,resonator,MeyerAPL,MeyerSSC,PootArxiv,BalandinNanoLetters}, which provided valuable
information about its microscale properties such as long-range
crystal order and inherent rippling \cite{MeyerSSC}. Graphene
membranes with lateral dimensions of the order of 0.1--1 $\upmu$m
were previously fabricated either by etching a substrate material
away from beneath a graphene crystallite, which left it supported by
a gold `scaffold' structure \cite{suspend}; by direct transfer of
graphene crystals onto an amorphous carbon film \cite{MeyerAPL}, or
by cleavage on silicon wafers with etched trenches \cite{resonator,PootArxiv,BalandinNanoLetters}.
The small sample size, especially for the case of suspended
graphene, remains a major limiting factor in various studies and
precludes many otherwise feasible experiments.

\begin{figure}
\begin{center}
\includegraphics[width=8.5cm]{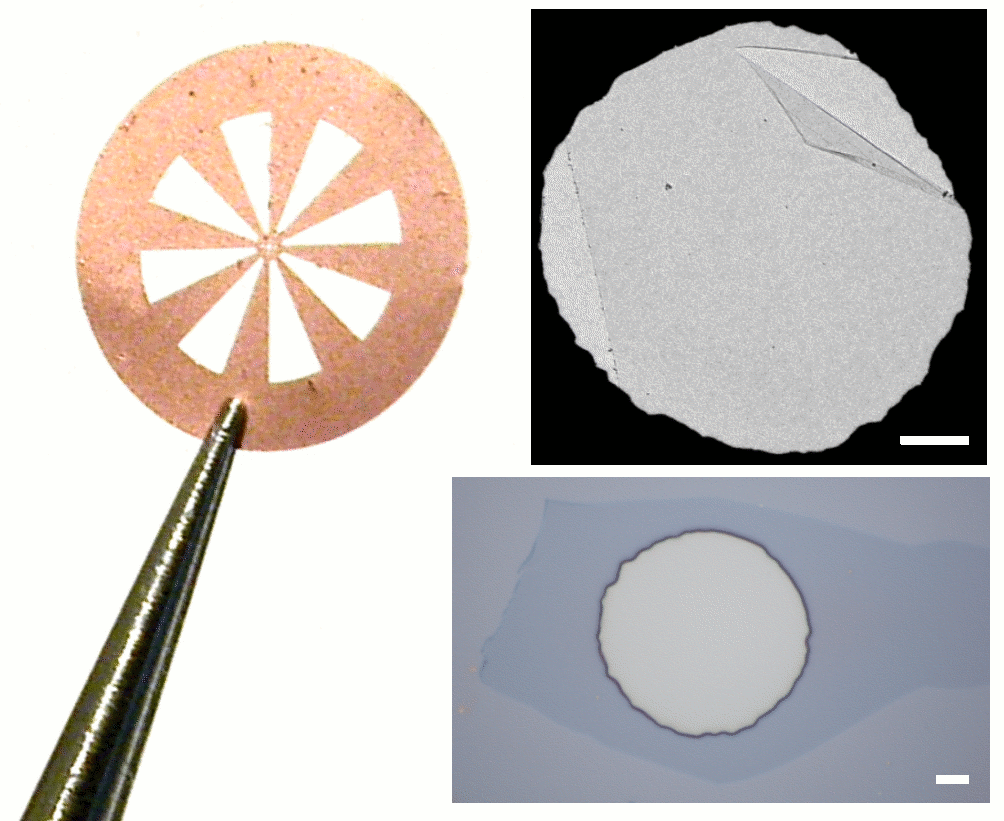}
\caption{\label{fig:grid} Graphene membranes. Left: Photograph of a
standard support grid for TEM (3~mm in diameter) with a central
aperture of 50~$\upmu$m diameter covered by graphene. Bottom:
Optical image of a large graphene crystal covered by photoresist in
the place where the aperture is planned. Top: TEM micrograph of one
of our graphene membranes that was partially broken during
processing, which made graphene visible in TEM. Scale bars:
5~$\upmu$m.}
\end{center}
\end{figure}


In this communication we report a technique for making large graphene
membranes with sizes that are limited only by the size of initial
flakes obtained by micromechanical cleavage, currently up to 100
$\upmu$m diameter. These membranes can be produced reliably from
chosen crystallites with a typical yield of more than 50\%. The
final samples are mechanically robust, easy to handle and compatible
with the standard holders for transmission electron microscopy
(TEM), which allows the use of graphene as an ultimately thin and
non-obstructing support in electron diffraction or high-resolution
transmission electron microscopy studies (see Fig. \ref{fig:grid}).
Furthermore, our procedures do not involve any aggressive etchants
that can lead to the `oxidation' of graphene~\cite{Ruoff} and/or its
irreversible contamination, which makes the technique suitable for
incorporation into complex microfabrication pathways. The membranes
demonstrated here should facilitate further studies of mechanical,
structural, thermal, electrical and optical properties of this new
material because graphene samples can now be used in a much wider
range of experimental systems. We have also found that graphene does
not meet the current perception of these one-atom-thick films as
being extremely fragile and prone to folding and
scrolling\cite{Nanoscroll2,Nanoscroll}. In fact, graphene appears to
be so stiff and robust that crystallites supported by one side can
freely extend ten microns away from a scaffold structure. The latter
observation is explained within elasticity theory by a huge Young's
modulus of graphene.

\begin{figure}
\begin{center}
\includegraphics[width=8.5cm]{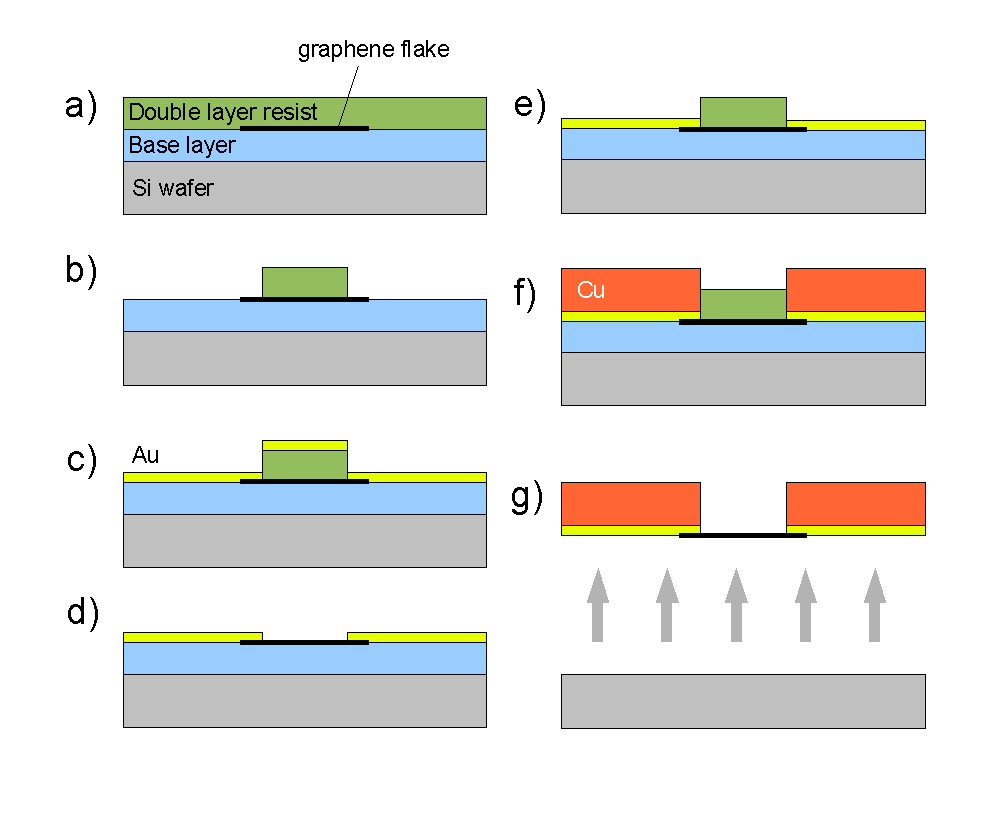}
\caption{\label{fig:method}Microfabrication steps used in the
production of graphene membranes.}
\end{center}
\end{figure}

Figure 1 shows examples of our final samples whereas Fig. 2 explains
the fabrication steps involved. Graphene crystals are first prepared
by standard micromechanical cleavage techniques\cite{aPNAS}. Sufficiently
large flakes produced in this way are widely distributed over a
substrate (occurring with a typical number density of $<1$ per
cm$^{2}$) and in a great minority as compared to thicker flakes.
This prevents their identification via atomic-resolution techniques
such as scanning probe or electron microscopies either due to
prohibitively small search areas or a lack of response specific to
single-layer graphene\cite{aPNAS}. Fortunately, one-atom-thick
crystals can still be identified on surfaces covered with thin
dielectric films due to a color shift induced by graphene, which
allows crystals to be found rapidly with a trained eye and a quality
optical microscope\cite{visible}. In the current work, we have used
Si wafers that, in contrast to the standard approach
\cite{aEFE,aPNAS,zAFMgraphene,PootArxiv}, are not oxidized but instead covered
with a 90~nm thick film of polymethyl methacrylate (PMMA) (referred
to as a base layer in fig. \ref{fig:method}-a). The optical
properties of PMMA are close to those of SiO$_2$, and the visible
contrast of graphene is optimal at this particular
thickness\cite{visible}. The PMMA film also serves later as a
sacrificial layer during the final liftoff (see below).

Once a suitable graphene crystal is identified in an optical
microscope, we employ photolithography to produce a chosen pattern
(in our case, a TEM grid) on top of graphene (we usually used a
double-layer resist consisting of 200~nm polymethyl glutarimide
(PMGI) from $MicroChem~Corp$ and 200~nm S1805 from
$Rohm~and~Haas$)(Fig. \ref{fig:method}-a,b). A 100~nm Au film with a
5~nm Cr adhesion layer is thermally evaporated after developing the
resist (Fig. \ref{fig:method}-c). Liftoff of the metal film is not
performed in acetone, which would destroy the base layer, but in a
2.45 wt \% TMAH solution (MF-319 developer; $MicroChem$) at
70$^{\circ}$C, resulting in a minimal etch rate for PMMA ($<5$\AA
min$^{-1}$)\cite{PMMAmask}(Fig. \ref{fig:method}-d).

The next step involves another round of photolithography (Fig.
\ref{fig:method}-e), in which the graphene crystal is remasked with
the same photoresist. The mask serves here to protect graphene
during electrodeposition, when a thick copper film is
electrochemically grown on top of the Au film, repeating the
designed pattern (Fig. \ref{fig:method}-f). We have chosen a
CuSO$_4$/H$_2$SO$_4$ electrolyte because of its low toxicity, resist
and substrate compatibility and ease of deposition. Finally, acetone
is used to strip the remaining resist, releasing the copper TEM grid
with the attached graphene membrane (Fig. \ref{fig:method}-g). The
sample is dried in a critical point dryer to prevent the membrane
rupturing due to surface tension. A copper thickness of 10-15~$\upmu$m
is found to be sufficiently robust for reliable handling of the samples. The
resulting membranes are then ready for transmission electron
microscopy and other graphene studies\cite{graphene137}.

\begin{figure}
\begin{center}
\includegraphics[width=8.5cm]{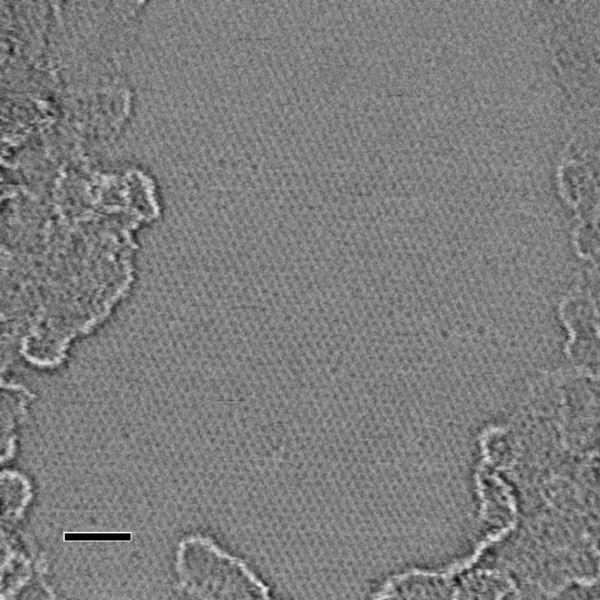}
\caption{\label{fig:lattice} High resolution bright field micrograph
of single-layer graphene. The image was taken at 100~keV with the
Daresbury SuperSTEM fitted with a Nion spherical aberration
corrector. Contamination is visible at the edges of the field.
Several dark spots seen within the clean central area are the
beam-induced knock-on damage that becomes increasingly more
pronounced for extended exposures. Scale bar: 2~nm.}
\end{center}
\end{figure}


Figure \ref{fig:lattice} shows an atomic-resolution TEM image of one
of our membranes. The crystal lattice of graphene is readily visible
in the clean central area of the micrograph, which is surrounded by
regions with hydrocarbon contamination. In the clean region, one can
also notice a number of defects induced by electron-beam exposure
(100~keV). Note that, prior to TEM studies, our membranes were
annealed in a hydrogen atmosphere at 250~$^\circ$C, which allowed
the removal of contaminants such as, for example, resist
residues\cite{STMclean}. Nevertheless, graphene is extremely
lipophilic, and we find that a thin contamination layer is rapidly
adsorbed on membranes after their exposure to air or a TEM vacuum.

Annealing the samples at temperatures higher than 300$^\circ$C is
found to trigger redeposition of copper and the formation of
nanoparticles on the surface of graphene (Fig.~4). These particles
are useful as a source of high contrast to aid focussing in TEM, and
as the in-situ calibration standard based on a copper lattice
constant. The top inset of Fig.~4 shows one such Cu crystal.
Furthermore, we have used the high angle annular dark field mode
(HAADF) of the SuperSTEM, which is very sensitive to chemical contrast.
Three foreign atoms found within one small area of a graphene
membrane are clearly seen on the HAADF image as white blurred spots
(lower inset of Fig. 4) and can be ascribed to adsorbed oxygen or
hydroxyl molecules. This illustrates that graphene membranes can be
used as an ideal support for atomically-resolved TEM studies.
Indeed, being one-atom-thick, monocrystalline and highly conductive,
graphene produces a very low background signal. Diffraction spots
due to graphene can be isolated and minimally obscure diffraction
patterns of investigated samples placed on such membranes. For
spectroscopic applications including x-ray microanalysis,
graphene also provides a minimal background due to the low atomic
number and a low concentration of impurities adsorbed on graphene's
surface.

\begin{figure}
\begin{center}
\includegraphics[width=8.5cm]{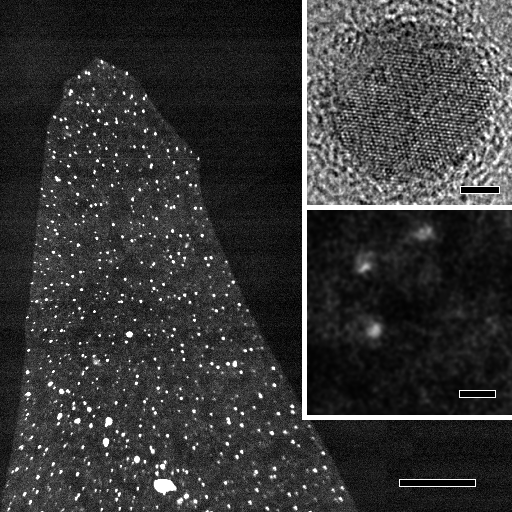}
\caption{\label{fig:beam} HAADF micrograph of a section of a
graphene membrane that fractured during annealing. The graphene
crystal is supported from one side only. White dots are copper
nanoparticles. Scale bar: 1$\upmu$m. Top inset: high resolution
bright field STEM micrograph of such a Cu particle ($\diameter$
8.0~nm; scale bar: 2~nm). Low inset: HAADF image of individual atoms
on graphene; scale bar: 2~\AA.}
\end{center}
\end{figure}

One of the most unexpected and counter-intuitive results of our work
is the observation of graphene crystallites supported from one side
only. Fig.~4 shows such a crystal left after a membrane was
fragmented during its annealing (probably due to thermal stress). In
this case, the graphene sliver extends nearly 10~$\upmu${m} from the
metal grid, in the absence of any external support. This contradicts
the perception that graphene is extremely supple and should curl or
scroll to minimize the excess energy due to free surface energy and
dangling bonds\cite{Nanoscroll2,Nanoscroll}. The previous
observations \cite{suspend,resonator,MeyerAPL} on suspended graphene
seemed to be in agreement with the latter assumption showing
scrolled edges\cite{suspend}. Figure \ref{fig:beam} proves that, on
the contrary, graphene is exceptionally stiff. We believe that the
fundamental difference between the case of Fig.~4 and the earlier
observations is that our crystals were fragmented in a gas
atmosphere rather than in liquid (our membranes broken in a liquid
were also strongly scrolled and folded).

To appreciate the stiffness of graphene, we note that the effective
thickness $a$ of single-layer graphene from the point of view of
elasticity theory \cite{LandauLifshitz} can be estimated as
$a=\sqrt{\kappa/E} \approx 0.23~$\AA, that is, smaller than even the
length of the carbon-carbon bond, $d=1.42~$\AA. Here we use the
bending rigidity $\kappa$ of $\approx 1.1~$eV at room
temperature~\cite{FasolinoNature}, and Young's modulus
$E\approx22$eV/\AA$^2$, which is estimated from the elastic modulus of
bulk graphite~\cite{moduli}. Therefore, the length $l$ of the
observed unsupported graphene beam is $\approx 10^6$ times larger
than its effective thickness. One could visualize this geometry as a
sheet of paper that extends 100~meters without a support. Even
though such extraordinary rigidity seems counterintuitive, it is in
good agreement with the elasticity theory as shown below.

Each carbon atom in the graphene lattice occupies an area
$S_0=\frac{3\sqrt{3}}{4}d^2$, and graphene's density is given by
$\rho=M/S_0\cong7.6\cdot10^{-7}$kgm$^{-2}$, where $M$ is the mass of
a carbon atom. Let us first consider the simplest case of a
horizontal rectangular sheet of width $w$ and length $l$ that is
infinitely thin, anchored by its short side ($y$-axis) and free to
bend under gravity $g$. The total energy of the  sheet is given by
\begin{equation}
\Sigma
=\frac{\kappa}{2}w\int^{l}_{0}dx\left(\frac{d^2h}{dx^2}\right)^2-\rho{g}w\int^{l}_{0}dxh
\end{equation}
where $x$ is the distance from the anchor point at $x=0$, and $h(x)$
is the deviation from the horizontal axis which is uniform along
$y$. The solution that minimizes the energy and satisfies the
boundary conditions is (cf. Ref.~\cite{LandauLifshitz})
\begin{equation}
h(x)=\frac{\gamma{l^2}x^2}{4}-\frac{\gamma{lx^3}}{6}+\frac{\gamma{x^4}}{24},
\end{equation}
where $\gamma=\rho{g}/\kappa\approx0.5\cdot10^{14}$m$^{-3}$,
$g\rho\cong7.48\cdot10^{-6}$Nm$^{-2}$. This yields the maximum
bending angle $(dh/dx)_{x=l}=\gamma{l}^3/6$ and, for the membrane in
Fig.~4 ($l \approx 20~\upmu$m), implies bending angles of several
degrees.

The above expression is a gross overestimate for bending of real
graphene beams with $w \approx l$ because the discussed purely
one-dimensional case takes into account only the bending rigidity
and neglects in-plane stresses that inevitably appear in a
non-rectangular geometry in order to satisfy boundary
conditions~\cite{LandauLifshitz}. Indeed, sheets of an arbitrary
shape should generally experience two-dimensional deformations
$h=h(x,y)$ and, in the case of graphene, bending becomes limited
by the extremely high in-plane stiffness described by $E$. This
makes graphene beams much harder to bend because their
apparent rigidity becomes determined by stretching rather than
simple bending. Elasticity theory provides an estimate for the
typical out-of-plane deformation $\bar{h}$ (see chapter 14 in ref. 18)
\begin{equation}
\frac{\bar{h}}{l}\approx\left(\frac{\rho{g}l}{E}\right)^{1/3}\approx(3\cdot
10^{-14}l)^{1/3},
\end{equation}
where $l \approx w$ is expressed in micrometers. This means that the
gravity induced bending is only of the order of 10$^{-4}$ for
graphene slivers such as shown in Fig.~4. We can also estimate the
corresponding in-plain strain as $(\bar{h}/{l})^2 \approx10^{-8}$.
Note that the crystal also supports an additional weight of many Cu nanoparticles. We have estimated their average weight
density as being 1000 times larger that that of graphene itself.
This should result in 100 times larger strain but still of only
10$^{-6}$. Graphene is known\cite{elasticityCN} to sustain strain of
up to 10\% without plastic deformations, albeit edge defects can
reduce the limit significantly allowing for the local generation of
defects. Still, for the membrane in Fig.~4 to collapse it
would require an acceleration of the order of 10$^{6}g$. This shows
that one-atom-thick graphene crystals of a nearly macroscopic size
have sufficient rigidity to support not only their own weight but
significant extra loads and survive accidental shocks during
handling and transportation.

In addition to their intrinsic stiffness, graphene crystals are
often corrugated, which further increases their effective thickness
and rigidity. Microscopic corrugations (ripples) were previously
reported for suspended graphene\cite{suspend,MeyerSSC}. Some (but
not all) of our membranes also exhibited macroscopic corrugations,
which extended over distances of many microns and were probably
induced by accidental bending of the supporting grid or mechanical
strain during microfabrication. Similar to the case of corrugated
paper, the observed corrugations of graphene should increase its
effective rigidity by a factor $(H/a)^2$ where $H$ is the
characteristic height of corrugations \cite{corrug1,corrug2}. The
increase due to ripples is minor but can be dramatic in the case of
large-scale corrugations.


Finally, we note that the described technique for making large
graphene membranes can also be applied to many other two-dimensional
crystals\cite{aPNAS} and ultra-thin films, including those materials
that cannot withstand aggressive media (e.g., dichalcogenides). One
can also use the technique in the case of graphene grown epitaxially
on metallic substrates \cite{EpiIr,EpiRu} in order to either make
membranes or study and characterise the epitaxial material further.
In this case, the final step in Fig.~2 can be substituted by etching
away the substrate or peeling off the electrodeposited TEM grid.

In conclusion, we have demonstrated a technique for producing large
graphene membranes in a comparatively robust and integratable
format. These membranes present a qualitatively new kind of sample
support for TEM studies. More generally, large scale suspended
graphene samples should allow a wider range of characterization
techniques to be employed and will facilitate the incorporation of
graphene in various microelectronic, optical, thermal or mechanical
devices. This is a key enabling step for both the investigation and
technological development of this exciting new material. The
observed counter-intuitively high rigidity of graphene should change
our perception of this one-atom-thick material as fragile and
mechanically unstable.  It already allows us to understand the previously unexplained fact that graphene does not scroll\cite{Nanoscroll2,Nanoscroll} and can be deposited as flat crystals even after being dispersed in a liquid\cite{aEFE}.

We thank the Engineering and Physical Sciences Research Council (UK)
and the Royal Society.

\end{document}